\documentclass[12pt,preprint]{aastex}
\notetoeditor{refer to: Fiamma Capitanio. Institute: IASF, CNR-INAF,Via Fosso del Cavaliere, 100, 00133 Rome ITALY. Tel: +390649934557. E-mail: fiamma@rm.iasf.cnr.it}

\newcommand{\xte}{{\it RXTE}}

\newcommand{\nh}{N_{\rm H}}
\shorttitle{IGR J17464-3213 temporal and spectral behaviour}
\shortauthors{Capitanio F. et al.}
\begin{document}
\title{3-200 keV spectral states and variability of the INTEGRAL Black Hole binary IGR J17464-3213\altaffilmark{1}}
\author{F. Capitanio\altaffilmark{2}, P. Ubertini\altaffilmark{2}, A. Bazzano\altaffilmark{2},
P. Kretschmar\altaffilmark{3,4},A. A. Zdziarski\altaffilmark{5}, A. Joinet\altaffilmark{6}, E.J. Barlow\altaffilmark{7},
A.J.Bird\altaffilmark{7}, A.J.Dean\altaffilmark{7}, E. Jourdain\altaffilmark{6},
 G. De Cesare\altaffilmark{2}, M. Del Santo\altaffilmark{2}, L. Natalucci\altaffilmark{2},
 M.Cadolle Bel\altaffilmark{8}, A. Goldwurm\altaffilmark{8}}
\altaffiltext{1}{Based on observations with {\it INTEGRAL}, an ESA project with instruments and
science data centre
funded by ESA member states (especially the PI countries: Denmark,
France,Germany, Italy, Switzerland, Spain), Czech Republic and 
Poland and with the participation of Russia and the USA.}
\altaffiltext{2}{IASF, CNR-INAF, Rome, Italy, Via Fosso del Cavaliere 
100, I-00133 Rome, Italy}
\altaffiltext{3}{MPE, Postfach 1603, 85740 Garching, Germany}
\altaffiltext{4}{{\it INTEGRAL} Science Data Centre,Chemin d'Ecogia 
16,1291, Versoix, Switzerland}
\altaffiltext{5}{Centrum Astronomiczne im.\ M. Kopernika, 18 00716 
Warszawa, Poland}
\altaffiltext{6}{CESR,Toulouse, France}
\altaffiltext{7}{School of Physics and Astronomy, University of
Southampton,Highfield, Southampton, SO 17 1BJ, UK.}
\altaffiltext{8}{Sap-CEA, Saclay, F-91191 Gif-sur-Yvette, France}
\begin{abstract}
On March 2003, IBIS, the gamma-ray imager on board the
{\it INTEGRAL} satellite, detected an outburst from a new source, IGR
J17464-3213, that turned out to be a {\it HEAO-1} transient, H1743-322.
In this paper we report on the high energy behaviour of this BHC
 studied with the three main instruments onboard {\it INTEGRAL}.
The data, collected with unprecedented sensitivity in the hard X-Ray range,
 show a quite hard Comptonised emission from 3 keV up to 150
keV during the rising part of the source outburst, with no thermal
emission detectable. A few days later, a prominent soft disk multicolour
component appears, with the hard tail luminosity almost unchanged:
$\sim\!5\times 10^{-9}$\,erg\,cm$^{-2}$\,s$^{-1}$.
 Two months later, during a second monitoring
campaign near the end of the outburst, the observed disk component
was unchanged. Conversely, the Comptonised emission from the central-hot
part of the disk reduced by a factor of $\sim10$. 
We present here its long term behaviour in different energy ranges and the
combined JEM-X, SPI and IBIS wide band spectral evolution of this source.
\end{abstract}
\section{Introduction}
On 2003 March 21 during the Galactic Centre Deep Exposure (GCDE) Program {\it INTEGRAL} 
detected  a relatively bright source ($\sim$60 mCrab at 15-40 keV) named IGR J17464-3213
\citep{REV}. The source was then localised at R.A.\ (2000) $=17^{\rm h}46.3^{\rm m}$, Dec.\ $=-32\degr 
14.4^\prime$, with an error
box of 1.6 arcmin (90\% confidence) and  associated with
H1743-322 \citep{Mark}, a bright BHC observed by HEAO1 in 1977 with an intensity of 700 mCrab at
2-10 keV and localised with two possible 
positions \citep{Dox}. The error box of IGR J17464-3213 was compatible with only one
of these two, thereby resolving this 25 year old ambiguity. After the 2003 outburst, follow up
observations with $\xte$ and {\it INTEGRAL} 
reported strong flux and possible spectral variability \citep{Parmar}. 
 $\xte$ observed the source for 
the first time on 2003 March 29 during a PCA Galactic bulge scan followed by 
a pointed observation. The mean PCA fluxes, in the bands 2-10, 15-40 and 
40-100 keV, were 50, 200, 220 mCrab respectively. 
The spectrum was consistent with an absorbed power law with a photon index 
1.49$\pm$ 0.02, and a column density:$\nh$= $2.4\times10^{22}\ cm^{-2}$ \citep{Mark}. 
In this paper we focus on the data collected with the coded mask co-aligned telescopes
IBIS (15 keV - 10MeV)\citep{Uber}, SPI (20 keV - 8 MeV)\citep{Ved} and JEM-X (3-35 keV)\citep{Lund} 
on board {\it INTEGRAL} \citep{Winkl}. 
The source was monitored by {\it INTEGRAL} in the framework of the Core 
Programme (CP) observations \citep{Kre,Gre}.
The data analysed here cover the period from the beginning of the outburst 
(March 2003) to its end (October 2003).
\section{Data Analysis}
%
The analysed data set consists of all CP observations in which
IGR J17464-3213 was within the high energy detectors field of
view. In particular, the observations were performed in three different
periods: from 2003 March 12 to 2003 April 15, from 2003 August 19 to 20, from
2003 October 4 to 9. The first period contains four observations of the pre-outburst state of the source; 
the other two shorter observations cover the end of the outburst phase.
All the CP observations are organised into un-interrupted 2000s long science windows (SCW):
light curves, hardness ratio and spectra are then extracted
for each individual SCW. 
Wide band spectra of the source are obtained using data from the three 
high energy instruments (JEM-X, IBIS and SPI).
In particular, the IBIS data were processed using the Off-line Scientific 
Analysis (OSA version 3) \citep{Gold} released by the {\it INTEGRAL} Scientific Data Centre 
(ISDC) \citep{Courv} with a modified pipeline in order to optimise the source spectral extraction.
The IBIS data set has been divided in two subsets according to
the source position being within the partially coded field of view (PCFOV)
or in the narrower fully coded field of view (FCFOV). Both data sets
have been used, after source intensity correction for off-axis
detectors response, to produce the source light curves in different energy
ranges. The off-axis correction was made using the available calibration
data, the accuracy on the flux being from 10\% at low energy (15-20 keV) up to 5\% at higher 
energy ($>$ 60keV). Only FCFOV data, less affected by the off-axis response of the gamma-ray 
instruments, was selected for spectral extraction and fitting.\\
SPI spectra were extracted using software specially developed for fitting the positions 
and the fluxes of all significant sources in the field of view. The background model used for SPI data is
based on an uniformity map determined for each energy band from empty fields observation.\\
JEM-X spectra were derived with the ISDC OSA version 4 release, and updated spectral matrices.
While IBIS and SPI provide a very large FOV ($>$30 deg), Jem-X has a narrower FOV ($>$10 deg), thus providing 
only a partial overlap with the high energy detectors. 
When spectral data are obtained in the same time interval from more then
one instrument, the proper normalizing constant is added in the data fit, taking into
account the best knowledge of the instrument cross calibration. In optimum conditions,
the {\it INTEGRAL} spectra we produced cover the range from 3 keV to 200 keV.
\section{Results}
\subsection{The time evolution of the outburst}
\begin{figure}
\begin{center}
\includegraphics[angle=0, scale=0.55]{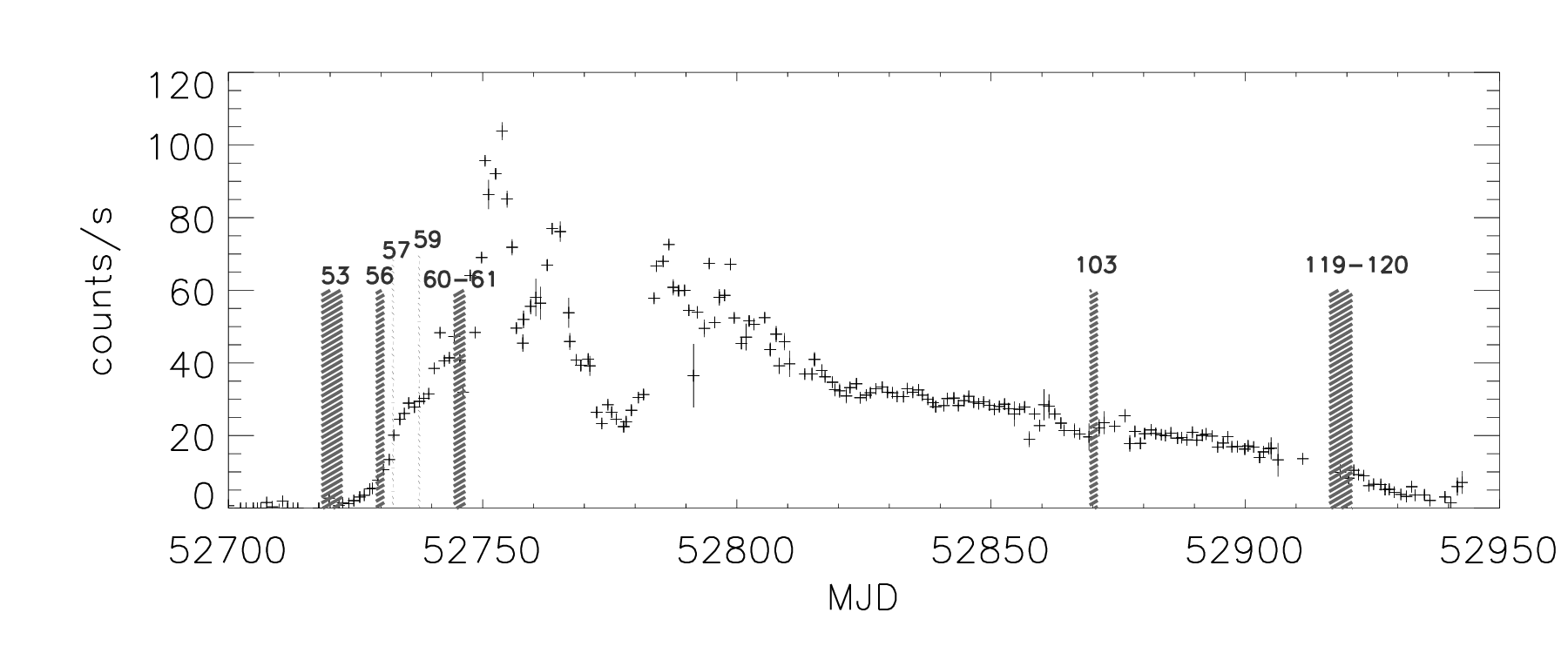}
\end{center}
\vspace{-0.6cm}
\caption{$\xte$-ASM one day averaged light curve. The gray dashed rectangles represent the simultaneous {\it INTEGRAL} observation periods corresponding to the spectra in figure~\ref{fig:fig4}. The time scale is
represented in Modified Julian Date}
\label{fig:fig1}
\end{figure}

\begin{figure}
\begin{center}
\includegraphics[scale=0.6]{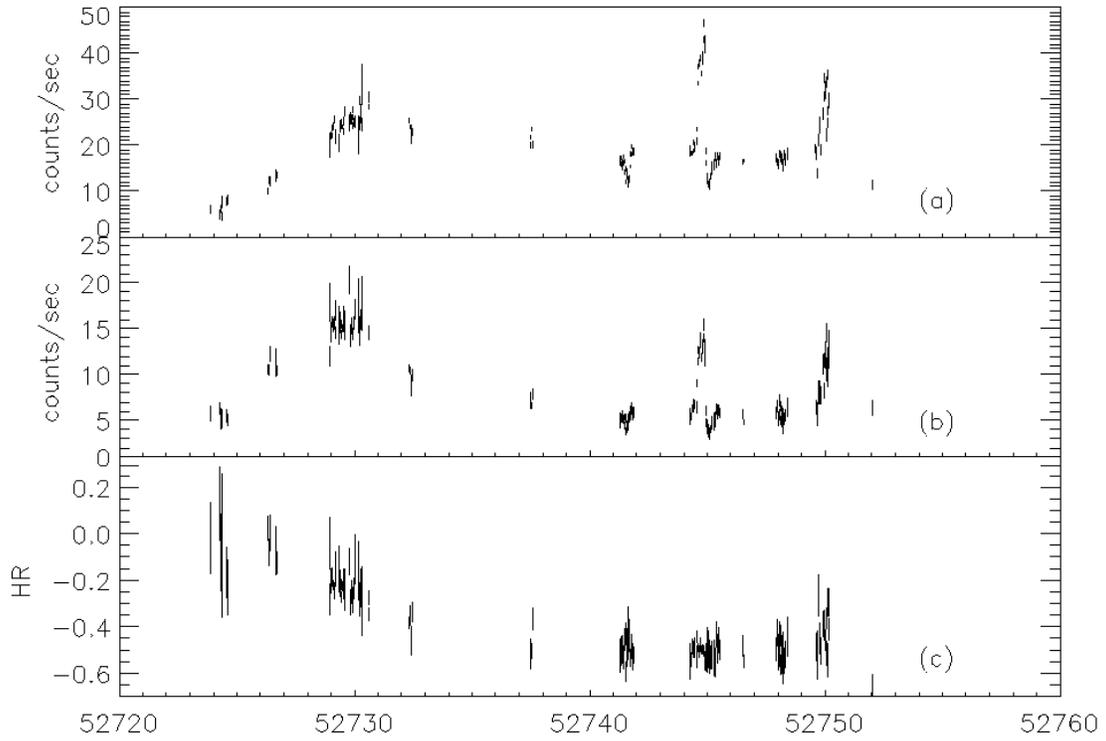}
\end{center}
\vspace{-0.6cm}
\caption{(a) 20-30 keV and (b) 40-60 keV IBIS light curves from revolution 53 to revolution 61. The hardness ratio derived from the two energy ranges is shown in the bottom part of the figure (c)}
\label{fig:fig2}
\end{figure}

\begin{figure}
\includegraphics[angle=90, scale=0.6]{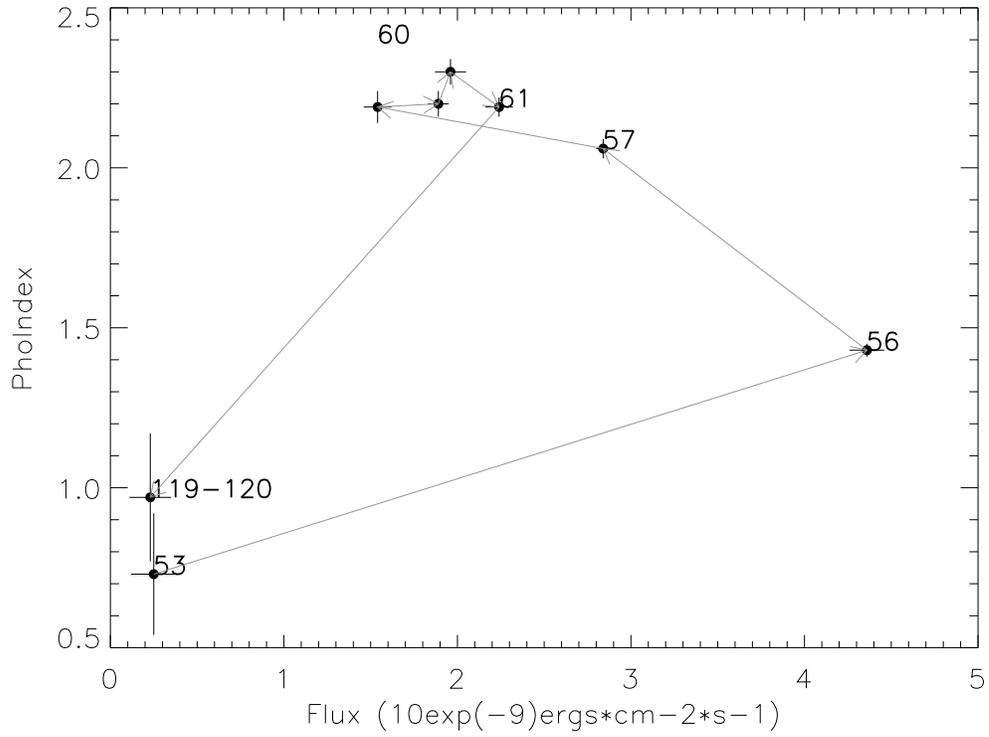}
\caption{Flux vs power law photon index for every SCW. The labels indicate the revolution number.}
\vspace{-0.8cm}
\label{fig:fig3}
\end{figure}

The source was continuously monitored by the $\xte$ ASM and whenever possible by {\it INTEGRAL} as shown in
Figure~\ref{fig:fig1}, with the simultaneous {\it INTEGRAL} observation periods 
represented by dashed rectangles. The monitoring of the source was quite continuous during the pre-outburst episode (from revolution 53 to 61) and
then only two deep observations were performed after about 4 (revolution 103) 
and 5.5 month (revolutions 119-120). In our work we have not analysed the 
data of the outburst itself as they are proprietary data published elsewhere \citep{Parmar}. 
We have produced light curves in the 20-30, 30-40, 40-80 keV energy ranges and investigated the Hardness Ratio, 
defined as $HR=({flux_{40-80}-flux_{20-30})/(flux_{40-80}+flux_{20-30}})$.
The IBIS 20-30 keV and 40-60 keV light curves of the initial part of the outburst, 
flux averaged over one SCW, are shown in Figure~\ref{fig:fig2}.
The IBIS light curves show strong variations and it is possible to note several short peaks in the flux 
intensity with different shapes in the lower and higher energy bands.
From the beginning of the IBIS observation, MJD $\sim$52729 to
$\sim$52731, (Modified Julian Date) the source averaged intensity increases monotonically
in the lower energy range while the flux is almost unchanged at higher energies: this
is reflected in the anticorrelation shown by the HR that decreases
continously in the same week as can be seen in Figure~\ref{fig:fig2}(c). 
The few points sampled in the next 20 days,
due to a limited coverage, show a decrease in both low energy flux and  HR. 
Starting from day $\sim$52744, the source shows a clearly visible flux increase in a few hours by a factor of 2.6, 3.0 
and 3.2 in the 20-30, 30-40 and 40-60 keV ranges respectively, without any corresponding
hardening. After the peak, the source
emission falls by a factor of $\sim$4.5  in the following hours
(20-30 keV) with a small decrease of the HR; the soft part of the spectrum becomes now more
prominent (Figure~\ref{fig:fig2}). This
corresponds to a dip in the flux emission, lasting a few hours (around day
52745.1), during which the source shows the lowest observed value for the HR.
Finally, the high energy component, and the HR, increase again
(MJD=52745.5). The source shows a similar behaviour around MJD=52750.
To look for a possible correlation of the Flux vs. Spectral index we have fitted the
available IBIS spectra during revolution 53, 57, 59-61 and 119-120, 
with a Cutoff power law model. We noticed that the cut off energy of the model had 
substantially the same value in each fit, so we freeze it to an average value of E=53.3 keV. The reduced spectral fit
$\chi^2$ covers a range from 0.8 to 1.3. 
The result, summarised in Figure~\ref{fig:fig3}, shows that the spectral index has a ``circular" or ``hysteresis" like
behaviour. In fact, at the end of the main outburst (revolution 103-120), the
source goes back in a hard state, very similar to the one of the pre-outburst IBIS measurements 
(revolution 53), but in the lower energy range a disk black body component is still present. 
This behaviour is not unusual for BHC even if the physical reason is not fully understood \citep{Zias}.
\subsection{Spectral evolution of the source}
The whole set of data has been fitted with standard XSPEC tools and the
best fits have been obtained with a multicolor disk black body (Diskbb) model \citep{Mit}
 for the low energy data and a Termal Comptonization model (CompTT) \citep{Tit} 
for the non thermal component. CompTT generally provides lower values for the plasma temperature 
$kT_{e}$ if compared with other models (i.e. Compps) while modelling hard state spectra of 
BH transients, the advantage being the proper modelling of plasma with high optical thickness.
The detailed spectral behaviour of the source is shown in Figure~\ref{fig:fig4}.
At the beginning of the outburst (Rev53) the source is very weak and therefore the flux in the lower part of the JEM-X spectrum is below the instruments sensitivity.
Therefore, it was possible to extract only a few points from JEM-X data in the range 15-22 keV and only a 
2$\sigma$ upper limit at lower energy. 
A single CompTT model fits the whole data set (15-150 keV) well, as shown in Table 1 with a $\textit{kT}$$_\textit{e}$$\sim$ 20 keV and a $\tau_{p}$$\sim$ 3.
The point corresponding to revolution 53 in Figure~\ref{fig:fig3} shows the correlation between 
the photon index and the flux for this data, indicating a Low/Hard state
\citep{Gre}. In a few days the source flux is then increasing
quickly, and the statistics enable extraction of the SPI spectra as well (revolutions 56 and 60-61).
 The high energy spectral evolution clearly indicates that the source make a transition to the soft 
 state as also shown in the HR evolution (see Figure~\ref{fig:fig2}~(c)). 
During revolution 56 the source was out of the JEM-X FOV therefore, only in this case, there are no data for E$<$ 20 keV. 
But the source is quite luminous in the hard X-ray as shown by IBIS and, moreover, looking at the time 
spectral evolution (Figure~\ref{fig:fig4}), is possible to suppose that the source is still in a Low/Hard state. 
In  revolutions 57-59 the  3-200 keV data can be fitted with a cutoff power law or 
with a thermal Comptonization Model without adding any black body component (Table 1).
Conversely in revolutions 60-61 the disk black body appears very bright and the emission of this 
component account for 45$\%$ of the total luminosity. 
During the phase of decreasing flux (Rev 103), there was no evidence of a hard tail in the IBIS data (20-250 keV, 2 $\sigma$ upper limit corresponding to $0.2\times10^{9}$erg$\times$$cm^{-2}s^{-1}$) while the source was 
clearly detected by JEM-X (see figure~\ref{fig:fig4}) and the data fitted with a Diskbb component showing a large value of the normalization constant 
($\textit{N}$$_\textit{Disk}$, see Table 1), that is proportional to the square of the inner accretion disk radius \citep{Mit}. 
The presence of a Diskbb in IGR J17464-3213 soft state is confirmed by the $\xte$ observation
performed in 2003 May 28 (MJD=52787) \citep{Hom}.
Two months later, close to the end of the outburst (revolutions 119-120), most of the flux is still due to the disk
emission but a clear, though faint,comptonized high energy tail is present the IBIS data. The IBIS spectra taken at
revolution 53 and 119-120,(i.e. at the beginning and at end of the outburst) are compatible with the same CompTT
model parameters: it is clear that in both cases the source shows a hard tail in the high energy spectrum.
Moreover, at the end of the outburst a disk component is still present, as also confirmed by the $\xte$ 
analysis of the data collected immediately before this {\it INTEGRAL} observation (from 2003 August
26 to 2003 September 23) \citep{Luv}.
\section{Discussion and Conclusion}
\begin{figure}
\begin{center}
\vspace{-0.35 cm}
\includegraphics[angle=90, scale=0.8]{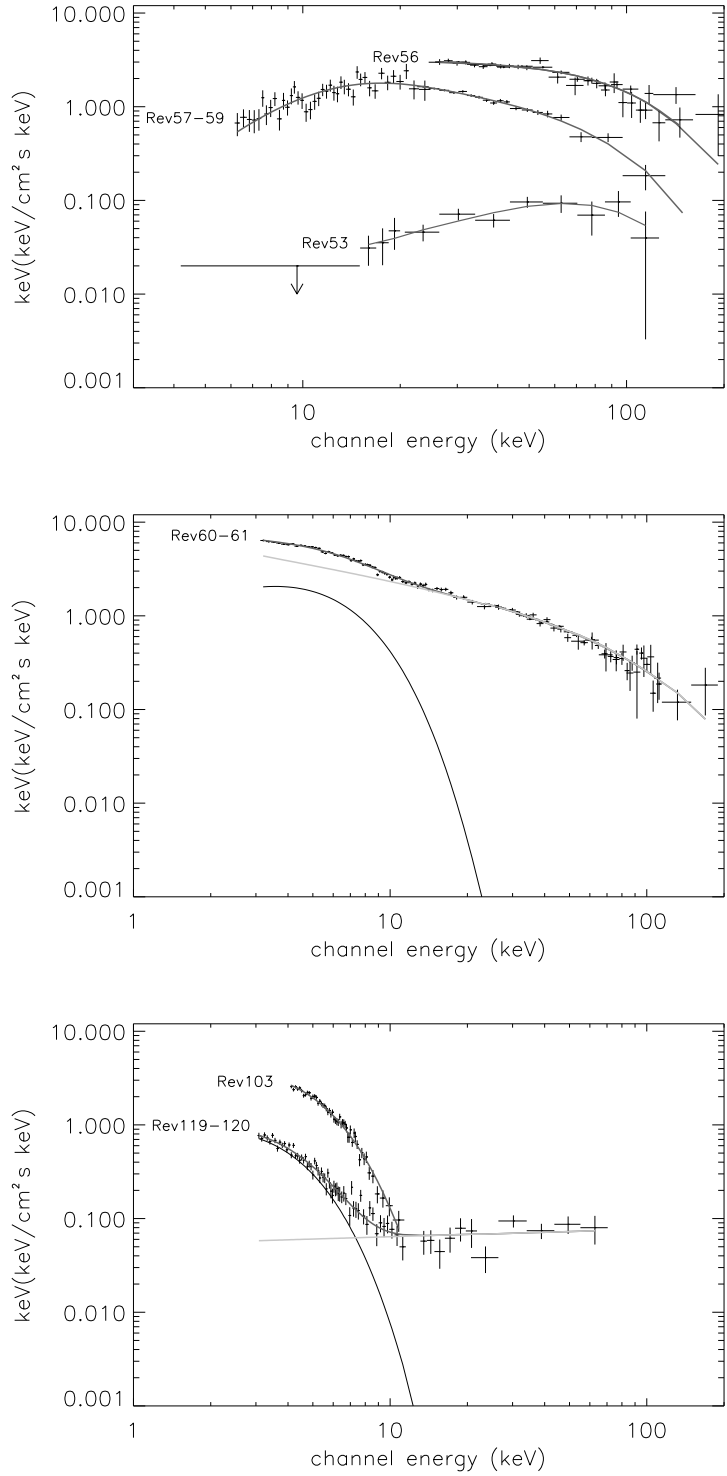}
\end{center}
\caption{IBIS JEM-X and SPI common spectra in different states of the source outburst:
the numbers on the top of each spectrum indicate the revolution number of the corresponding
{\it INTEGRAL} observation.}
\label{fig:fig4}
\end{figure}

The picture obtained from the spectral analysis in Figure~\ref{fig:fig4} is
quite clear, and summarised as follows.
Phase 1: The source is in a ``hard state". At the beginning of the {\it INTEGRAL}
monitoring campaign (revolution 53) the source has a low luminosity and the bulk of the
emission is in the energy range 50-90 keV via Comptonisation of soft
photons in an optically thick hot inner part of the disk or corona
($\textit{kT}$$_\textit{e}$$\sim$20
keV and $\tau_{p}$ $\sim$3). In this phase there is no evidence of a soft disk component emission,
with the disk being eventually truncated far from the hot inner region, though
injecting soft photons that are up-scattered by the inner hot flow. In the
next few days the amount of accreted matter is increasing dramatically and
in revolution 56 ($\sim$10 days later) the total emitted energy has increased by a
factor of 30: the temperature of the Comptonised plasma is not changed and
the optical thickness of the inner part of the corona become
smaller, i.e. from $\tau_{p}$$\sim$3 to $\tau_{p}$$\sim$1.3 (revolution
57-59). There is still no evidence of soft emission from the system. 
Slowly, the peak emission shifts
from 70 keV (revolution 53) to 40 keV (revolution 56) to finally 20 keV (revolution
57-59). The spectral evolution is shown in Figure~\ref{fig:fig4} (top panel).
Phase 2: The source is in a ``soft/intermediate state" (revolution 60-61). Over a
few days the average amount of energy released increases by a factor of 3 to 5, the
optical thickness of the inner part of the corona becomes lower and lower ($\tau_{p}$
$\sim$0.4),
 the temperature of the hot Comptonised part increases accordingly up
to $\sim$40 keV and a bright soft disk component become visible, being
responsible for the majority of the system energy release. The fit
parameters are compatible with a $T_{in}$ temperature of 1.5 keV and a disk
close to the last stable orbit. The spectrum obtained with JEM-X, 
IBIS and SPI (Figure~\ref{fig:fig4}, middle panel) 
has a high statistical significance showing the presence
of a substantial Comptonised component. 
A few days later the source has reached its maximum, showing a factor of two
more flux but basically unchanged high energy behaviour \citep{Parmar}.
Phase 3: The source enters the ``soft state". The last
observations were performed a few months after the main peak of the outburst
(revolution 103-120). The energy release has substantially decreased 
(factor of $\sim$3), the hard emission has basically switched off, and the
Comptonised photons energetics dropped by a factor of more than 10 with emission peaked at about
70 keV as in the pre-outburst state (revolution 53). This primarily indicates
the temperature of the disk dropping to 0.9 keV, and the disk substantially
outside from the last stable orbit. Soon afterwards, the source was no longer
detectable by {\it INTEGRAL}; one month later it 
was also below the ASM/$\xte$ detection capability.
The {\it INTEGRAL} observation of IGR J17464-3213 has been essential to
disentangle the source behaviour and its spectral state evolution and, in
turn, to better understand the physical processes active in the system in
the different states.
\acknowledgments
We acknowledge the ASI financial/programmatic support
via contracts I/R/389/02 and I/R/041/02; C. Spalletta for the careful editing of the manuscript, 
M. Federici for supervising the {\it INTEGRAL} data analysis
system and A. Tarana for useful scientific discussions.

\clearpage
\clearpage
\begin{center}
\begin{deluxetable}{lccccccccc}
\tabletypesize{\footnotesize} \tablenum{1} \tablewidth{0pt}
\tablecaption{Fit parameter values (90 \% confidence)} \tablehead{
\multicolumn{1}{l}{Revolution}& \multicolumn{1}{c}{$T_\textit{in}$
}& \multicolumn{1}{c}{$N_\textit{Disk}$}& \multicolumn{1}{c}{$\textit{kT}$$_\textit{e}$}&
\multicolumn{1}{c}{$\tau_{p}$}&
\multicolumn{1}{c}{$N_\textit{Comp}$ $\times10^{2}$}&
\multicolumn{1}{c}{$\chi^2_\nu$}&
\multicolumn{1}{c}{Deg.of}&
\multicolumn{1}{c}{flux $(\times10^{9}$}&
\multicolumn{1}{l}{Fluxband}
\\
\multicolumn{1}{l}{}&
\multicolumn{1}{c}{keV}&
 \multicolumn{1}{c}{}&
 \multicolumn{1}{c}{keV}&
\multicolumn{1}{c}{}&
 \multicolumn{1}{c}{}&
\multicolumn{1}{c}{}&
\multicolumn{1}{c}{freedom}&
 \multicolumn{1}{c}{erg$\times$$cm^{-2}s^{-1}$)}&
 \multicolumn{1}{l}{ keV}}
\startdata
Rev53&\nodata&\nodata&$20^{+15}_{-5}$&$3^{+4}_{-2}$&$2.38^{+0.01}_{-0.01}$&0.7&7&$\sim$0.3&(15-250)\\
\\
Rev56&\nodata&\nodata&$24^{+3}_{-3}$&$1.5^{+0.2}_{-0.3}$&$8.3^{+0.8}_{-0.7}$&0.96&14 &$\sim$7&(20-250)\\
\\
Rev57-59&\nodata&\nodata&$19^{+5}_{-3}$&$1.3^{+0.2}_{-0.4}$&$1.5^{+03}_{-0.3}$&1.03&57 & $\sim$6&(3-250) \\
\\
Rev60-61&$1.5^{+0.1}_{-0.1}$&$77^{+16}_{-18}$&$33^{+14}_{-8}$&$0.4^{+0.2}_{-0.2}$&$8.3^{-2}_{+2}$&1.01&106&$\sim$20&(3-250)\\
\\
Rev 103&$1.09^{+0.02}_{-0.02}$&$357^{-26}_{+39}$&\nodata&\nodata&\nodata&1.09&57&$\sim$3&(3-20)\\
\\
Rev119-120&$0.87^{+0.02}_{-0.02}$&$402^{+344}_{-214}$&$>$13&$<$3&$1.5^{+5}_{-0.3}$&1.10&68&$\sim$1&(3-250)\\
\enddata
\end{deluxetable}
\end{center}

\begin{thebibliography}{b}
\bibitem[Courvoisier et~al.(2003)]{Courv} 
Courvoisier T.J.L., R. Walter, V. Beckmann, et al., A\&A, 2003, 411,L53
\bibitem[Doxsey et~al.(1977)]{Dox} 
Doxsey, H. Bradt, G. Fabbiano, et al., 1977, IAU Circ.3113
\bibitem[Goldwurm et~al.(2003)]{Gold}
Goldwurm A., P. David, L. Foschini, et al., A\&A, 2003, 411, L223
\bibitem[Grebenev et~al.(2003)]{Gre}
Grebenev S. A., Lutovinov A. A., Sunyaev R. A. et al. 2003 Atel 189
\bibitem[Homan et~al.(2003)]{Hom}
Homan J., Miller, J.M., Wijinands, R, et al. 2003 Atel 162
\bibitem[Kaluzienski et~al.(1977)]{Kaluz}
Kaluzienski L.J. \& Holt S.S., 1977, IAU Circ.3099
\bibitem[Kretschmar et~al.(2003)]{Kre}
Kretschmar, P., Chenevez, J., Capitanio, F., et al. 2003 Atel 180
\bibitem[Lund et~al.(2003)]{Lund}
Lund N., C. Butz-Jorgentsen, N.J. Westergaard, et al., A\&A, 2003, 411,L231
\bibitem[Lutovinov et~al.(2004)]{Luv}
Lutovinov A., Revnivtsev M., S.Molkov et al. 2004 astro-ph/0407342
\bibitem[Markwardt et~al.(2003)]{Mark}
Markwardt, C.B. \& Swank,J.H. 2003, Atel 133
\bibitem[Mitsuda et~al.(1984)]{Mit}
Mitsuda K., Inoue H., Koyama K. et al. 1984, PASJ, 36, 741
\bibitem[Parmar et~al.(2003)]{Parmar} 
Parmar A.N., E. Kuulkers, T. Oosterbroek, et al., A\&A, 2003, 411, L421
\bibitem[Revnivtsev et~al.(2003)]{REV}
Revnivtsev M., Chernyakova, M., Capitanio, F., et al. 2003, Atel 132
\bibitem[Titarchuk et~al.(1994)]{Tit}
Titarchuk L. 1994 ApJ,434,570
\bibitem[Ubertini et~al.(2003)]{Uber}
Ubertini P., F. Lebrun, G. Di Cocco, et al., A\&A, 2003, 411, L131
\bibitem[Vedrenne et~al.(2003)]{Ved}
Vedrenne, G., J.P. Roques, V. Shoenfelder, A\&A 2003, 411, L63
\bibitem[Winkler et~al.(2001)]{Winkl}
Winkler C. \& T.L.L. Courvoisier, et al., A\&A, 2003, 411, L1.
\bibitem[A. Zdziarski et~al.(2004)]{Zias}
A. Zdziarski \& Gierlinski. Progress of theoretical Physiscs 2004 in press
\end{thebibliography}
\end{document}